\documentclass[letters,fleqn,usenatbib]{mnras}
\usepackage[fleqn]{amsmath}
\usepackage{graphicx}
\usepackage[varg]{txfonts}
\usepackage[T1]{fontenc}
\usepackage{ae,aecompl}

\title[Metallicity dependance of {[Y/Mg]} -- age relation]{On the metallicity dependance of the {[Y/Mg]} -- age relation for solar type stars}

\author[S.~Feltzing et al.]{Sofia~Feltzing,$^{1}$\thanks{{\tt email: sofia@astro.lu.se}} Louise~M.~Howes,$^{1}$  Paul~J.~McMillan,$^{1}$ and Edita~Stonkut{\.e}$^{1,2}$ \\
 $^{1}$Lund Observatory, Department of Astronomy and Theoretical Physics, Box 43, SE-22100, Lund, Sweden\\
$^{2}$Institute of Theoretical Physics and Astronomy, Vilnius University, Saul\.{e}tekio al. 3, LT-10222, Vilnius, Lithuania\\
}

\date{Accepted 2016 .Received 2016 ; in original form 2016 }

\pubyear{2016}

\begin{document}
\label{firstpage}
\pagerange{\pageref{firstpage}--\pageref{lastpage}}
\maketitle

\begin{abstract}
Several recent studies of Solar twins in the Solar neighbourhood have shown a tight correlation between various elemental abundances and age, in particular {[Y/Mg]}. If this relation is real and valid for other types of stars as 
well as elsewhere in the Galaxy it would provide a very powerful tool
to derive ages of stars without the need to resort to determining 
their masses (evolutionary stage) very precisely. The method would also likely work if the stellar parameters
have relatively large errors.  
The studies presented in the recent literature span a narrow range of {[Fe/H]}. By studying a larger sample of Solar neighbourhood dwarfs with a much larger range in [Fe/H], we find that the relation between {[Y/Mg]} and age depends on the 
{[Fe/H]} of the stars. Hence, it appears that the {[Y/Mg]} -- age relation is unique to Solar analogues. 
\end{abstract}


%--------------------------------------------------------------------

\section{Introduction}
\label{sec:introduction}

To build a full picture of how a galaxy like our own Milky Way formed and evolved we 
can make use of the properties of the stars (e.g., age and elemental abundances) as 
tracers of past events (e.g., the star formation history). Until recently the study of
elemental abundances and ages for stars was limited to the direct Solar neighbourhood
or special places in the Milky Way such as globular clusters or the bulge. This situation
is now rapidly changing with large spectroscopic data-sets \citep[e.g.,][]{Majewski15} coming
on-line and the first data-release from ESA's $Gaia$ satellite \citep[][]{2016arXiv160904172G,2016arXiv160904303L}. 

Determining elemental abundances is in principle straightforward, but not necessarily simple. See, for example \citet{Melendez14} for a comprehensive summary of  difficulties in the analysis of the Sun and solar analogues. Methods to enhance the precision in the abundance determinations
have been identified and resulted in some very interesting findings, such as the inhomogeneity
in elemental abundances found in the open cluster Hyades by \citet{Liu2016}.  Determining ages of stars is, on the other hand, truly difficult. The most commonly used  
method is to compare the absolute luminosity (or magnitude) and effective temperature (or colour) of the
star with theoretical models, isochrones or stellar evolutionary tracks. This 
requires that we have good measurements of  the apparent luminosity of the star,  the distance to the star, and either the temperature or the colour. In addition we 
need a  measure of the metallicity (or the iron content [Fe/H]) as
the isochrones and evolutionary tracks change as the metallicity changes \citep[e.g.,][]{Salasnich00}. In principle this method should be applicable to many types of stars -- however, in practise, it is currently limited to  turn-off stars  and stars on the sub-giant branch, as the isochrones are most sensitive to age in these regions of the Hertzsprung-Russell diagram. 
If we also have the mass we can place the star in the $T_{\rm eff}$--$\log g$ diagram together with the relevant isochrones and read off its age with much higher precision for many more evolutionary stages. Recently, asteroseismic data have made it possible to derive masses for stars on the red giant branch \citep{Chaplin13}. Regardless of the method used to place
the star in the Hertzprung-Russell diagram the uncertainty in the  age always depends strongly on the  errors in the measurements needed to place the star accurately on the correct stellar track.

\begin{figure*}
  \centering
    \includegraphics[width=8.5cm]{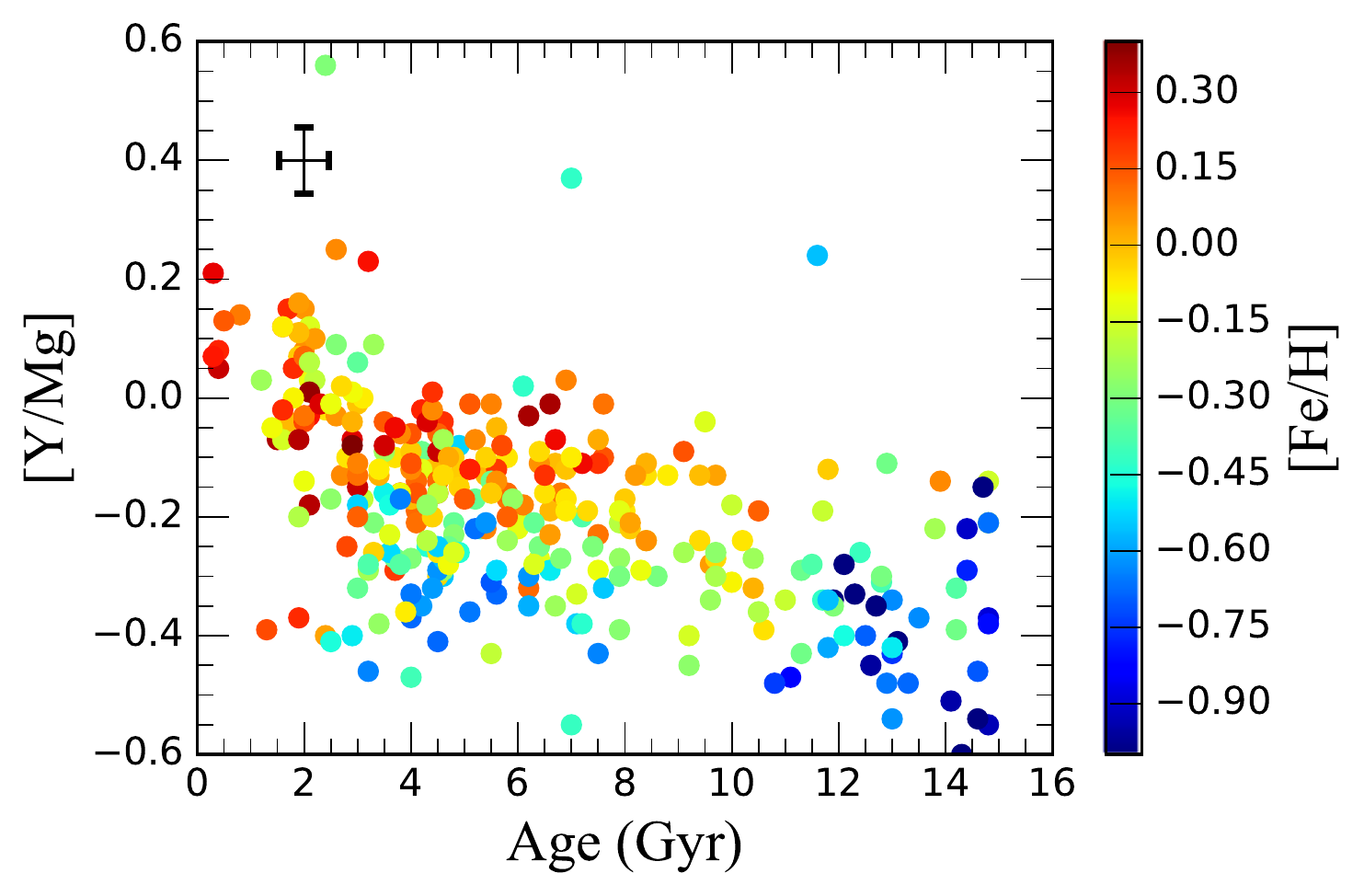}\includegraphics[width=8.5cm]{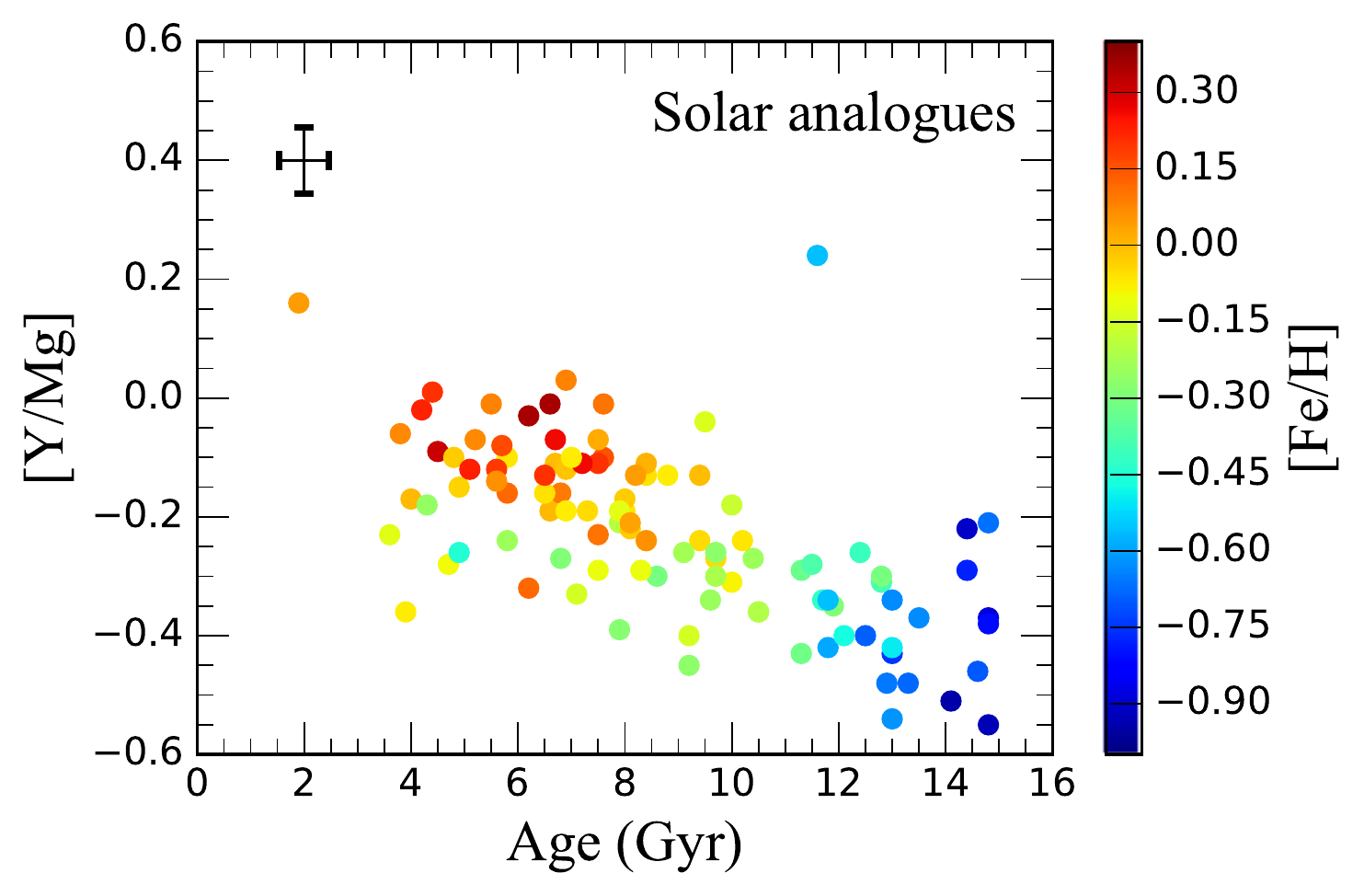}
  \vspace{-2mm}
      \caption{{\sl Left panel:} [Y/Mg] as a function of age (in Gyr) for all stars
      with $\sigma_{\rm Age}<$1\,Gyr from \citet{Bensby14}. {\sl Right panel:} [Y/Mg] as a function of age (in Gyr) for all stars
      with $\sigma_{\rm Age}<$1\,Gyr from \citet{Bensby14} that also have $T_{\rm eff}$ in the range $ 5777\pm 100$\,K (i.e. solar analogues). The stars
      are colour coded according to their [Fe/H] (as indicated in the colour bar). A typical error bar is shown.} 
         \label{fig:all}
  \vspace{-5mm}
\end{figure*}

The difficulties in obtaining good ages for all types of stars from isochrones 
and evolutionary tracks have fuelled an interest in alternative methods. We will
not discuss all of them here but refer to the comprehensive review by \cite{Soderblom10}. One new example includes studies that explore  mass indicators  in stellar spectra.
Using results from APOGEE \cite{Masseron15} show that the ratio of carbon to nitrogen ([C/N]) is an age indicator for red giant branch stars. Similarly, \citet{Ness16} and  \citet{Martig16}, using  different methods, find that NIR red giant spectra show features (mainly carbon lines) that are sensitive to mass. Other lines may also show a dependance on
mass. This has been little explored so far (mainly due to the lack of good mass determinations).  \citet{Bergemann16} show that the  H$\alpha$-line in K giants is 
sensitive to mass. Their finding is (so far) purely empirical.

Studies of elemental abundances in the Solar neighbourhood have shown some interesting
correlations. \citet{Bensby14} found that [Ti/Fe] correlates relatively
well with age. 
\citet{Battistini16} found several interesting
correlations between neutron capture elements, e.g., [Sr/Fe] and [Zr/Fe], and age.

The first evidence of a linear correlation between the ratio of yttrium to magnesium ([Y/Mg]) and stellar age was presented by \cite{daSilva12} for a sample of 25 solar-type stars. Recently, \cite{Nissen15} found a strong correlation between [Y/Mg] and age for a small sample of solar twin stars. This result has been reproduced by \citet{Spina16} and \citet{Tucci16}.  The total spread in [Fe/H] for the solar twins reported in these studies is about 0.3\,dex (about $\pm$0.15 above and below solar). In all three studies the very high precision is achieved thanks to a strictly differential analysis focusing on stars with very similar parameters. In most studies of elemental abundances in the Solar neighbourhood or elsewhere, this is not the case. Instead larger ranges of stellar parameters are included in order to obtain larger samples \citep[e.g.,][]{Edvardsson93}. 

 This is an intriguing result pointing to the possibility of using measurements of [Y/Mg] to derive the age of solar-like stars. The questions then naturally arise: How universal is this correlation? Does it hold for a larger range of [Fe/H]? Does it hold for stars not in the Solar neighbourhood? Does it hold for stars with different kinematics (i.e., thin vs. thick disc)? In this paper we aim to answer some of these questions. 

%--------------------------------------------------------------------

\section{Investigating the metallicity dependence of the [Y/Mg]--age relation}
\label{sect:investigation}

The main dataset used in this study is taken from \cite{Bensby14}. Briefly, \cite{Bensby14} analysed high-resolution, high signal-to-noise ratio spectra for 714 dwarf stars in the Solar neighbourhood. The stellar parameters are based on an analysis of the spectra, taking the 
stellar parallaxes from Hipparcos \citep{1997A&A...323L..49P} into account. Elemental abundances were obtained for several elements, including Mg. Additional elemental abundances for iron peak elements and some $r$- and
$s-$process elements have been presented in \citet{2015A&A...577A...9B} and \citet{Battistini16} \citep[but note that 
the Y-abundances are from][]{Bensby14}.  We refer the reader to the papers describing the full analysis for a discussion of how stellar parameters were
derived and the analysis of lines with, e.g., hyperfine structure was done.

The left panel of Fig.\,\ref{fig:all} shows the [Y/Mg] trend with age for all stars in \cite{Bensby14} that have $\sigma_{\rm Age}<$1\,Gyr. As in \cite{Nissen15} (their Fig. 10) we see a downward trend such that older stars have lower [Y/Mg]. The spread at a given age is significant and this spread results from stars with lower [Fe/H] on average having lower [Y/Mg] at a given age. 
The right panel of the figure shows the same plot but now only for stars that have $T_{\rm eff}$ $\pm $100\,K of that of the Sun, i.e., solar analogues. Solar twins are stars with $T_{\rm eff}$ and $\log g$ very close to those of the Sun \citep[][]{1978A&A....63..383H,1981A&A....94....1C}, whilst solar analogues is a more loose definition with stars being `very similar to the sun' but not necessarily a twin to the sun. The solar  analogues in Fig.\,\ref{fig:all} show the same downward trend of [Y/Mg] as age is increasing as found for the full sample. Also here we see that at a given age stars with higher [Fe/H] also have higher [Y/Mg]. This is particularly present for ages $4$-$8$\,Gyr. For the oldest stars the change of [Y/Mg] as a function of [Fe/H] is less obvious, but still noticeable.

We have done the same analysis for the dataset by \cite{2016ApJS..225...32B} of $\sim$1600 dwarf stars in the Solar neighbourhood. This sample consists of stars being monitored for radial velocity variations and hence potentially harbouring planets. The stars exhibit very similar trends in the [Y/Mg] versus age diagram as the stars in the \citet{Bensby14} sample, but as the sample stars are centered on solar metallicity and have very few stars even at --0.5\,dex in [Fe/H] the resulting trend with [Fe/H] is less obvious, although still present, giving further weight to our findings.

Are the trends we see in Fig.\,\ref{fig:all} influenced by the mixture of stellar populations present in the Solar neighbourhood (e.g., the thin and the thick disc) or are the findings universal? In the samples from \cite{Nissen15} and \cite{Spina16}  almost all the stars have kinematics very similar to that of the sun, what we refer to as thin disc kinematics and only a few stars in each sample have the enhanced $\alpha$-abundances typical of the  thick disk or thick disc kinematics. In
\citet{2016A&A...593A..65N} three stars with $\alpha$-enhanced abundances fit the thin disc trend for [Y/Mg] -- age well. This
is also the case for the ten stars with thick disk kinematics studied in \citet{Tucci16}. The sample by \citet{Bensby14} was chosen to probe the abundance trends in stars with typical thin and thick disc kinematics. When we divide the sample according to their kinematics into thin and thick disc stars \citep[following the probabilistic division used in][]{Bensby14} we find that the trends are a little less obvious but overall they persist.

The influence of `radial migration' in the Milky Way means that the stars in the Solar neighbourhood come from a variety of birthplaces, in a way that can not be simply determined from their kinematics \citep[e.g.][]{Roskar08,Schonrich09}. This implies that the [Y/Mg] trend observed locally does not necessarily apply to stars with a single birthplace, but to stars born across much of the Milky Way.  It  would therefore be valuable to determine observationally that the trend is seen elsewhere. Whilst waiting for such samples we can only resort to using the kinematics of the stars today to infer
their birth places by using, e.g., their mean distance to the Galactic centre as a proxy for their birth place \citep[compare, e.g.,][]{Edvardsson93}. Recently, \citet{2016A&A...592A..87A} used this for stars with solar like properties. In relation to the [Y/Mg] -- age relation they found no difference between stars from the 
inner disk and the solar neighbourhood, which is agreement with the results from \citet{Bensby14}.

As a check we have re-derived ages for the Bensby sample using a version of the Bayesian method described by \citet{Binneyetal14} with the Parsec isochrones \citep{ParsecIsochrones}. Our method differs from that of \citet{Binneyetal14} in that we have chosen to take a flat prior in both age and metallicity. We also take isochrones with a wider spread of metallicity and do not allow as wide a spacing in $Z$ between isochrones. We exclude the pre-main sequence phase from the isochrones, because there are no indications from the spectra that these stars are pre-main sequence. All trends with age persist with both methods. 

%--------------------------------------------------------------------

\section{Discussion}
\subsection{Potential effects of target selection} 

When we select stars for spectroscopic studies with high resolution and high signal-to-noise ratios there can be a tendency to select the brightest stars in the catalogue, as they will be the fastest to observe and we can thus get more stars in our sample. How this might affect the types of stars that make it into an `FGK-turn-off' sample has been explored, e.g., in \cite{Feltzing01} and \cite{Bergemann14}, where it is shown that old and metal-rich stars are preferentially disfavoured when a magnitude cut is imposed onto the sample. This is natural as old and metal-rich stars are both fainter and redder than young and metal-poor stars.

Can this type of selection effect have influenced the trend shown in Fig.\,\ref{fig:all}? It is possible, in so much as the trend of [Y/Mg] as a function of age for solar metallicity stars should be flatter (i.e., old and metal-rich stars should occupy high [Y/Mg] and old age, where there currently are no stars present). Do these types of stars exist? Several studies in the Solar neighbourhood have shown such stars but the evidence is not robust \citep[see, e.g.,][]{Feltzing01,Casagrande11}. 

We have also considered the possibility that the trends we see in this dataset are affected by the choice of [Fe/H]. At all but the oldest ages we have stars covering the full range of [Fe/H] seen in the discs of the Milky Way. Hence, we may conclude that this particular selection bias likely has not influenced our results. 

\begin{figure}
  \centering
    \includegraphics[width=8.5cm]{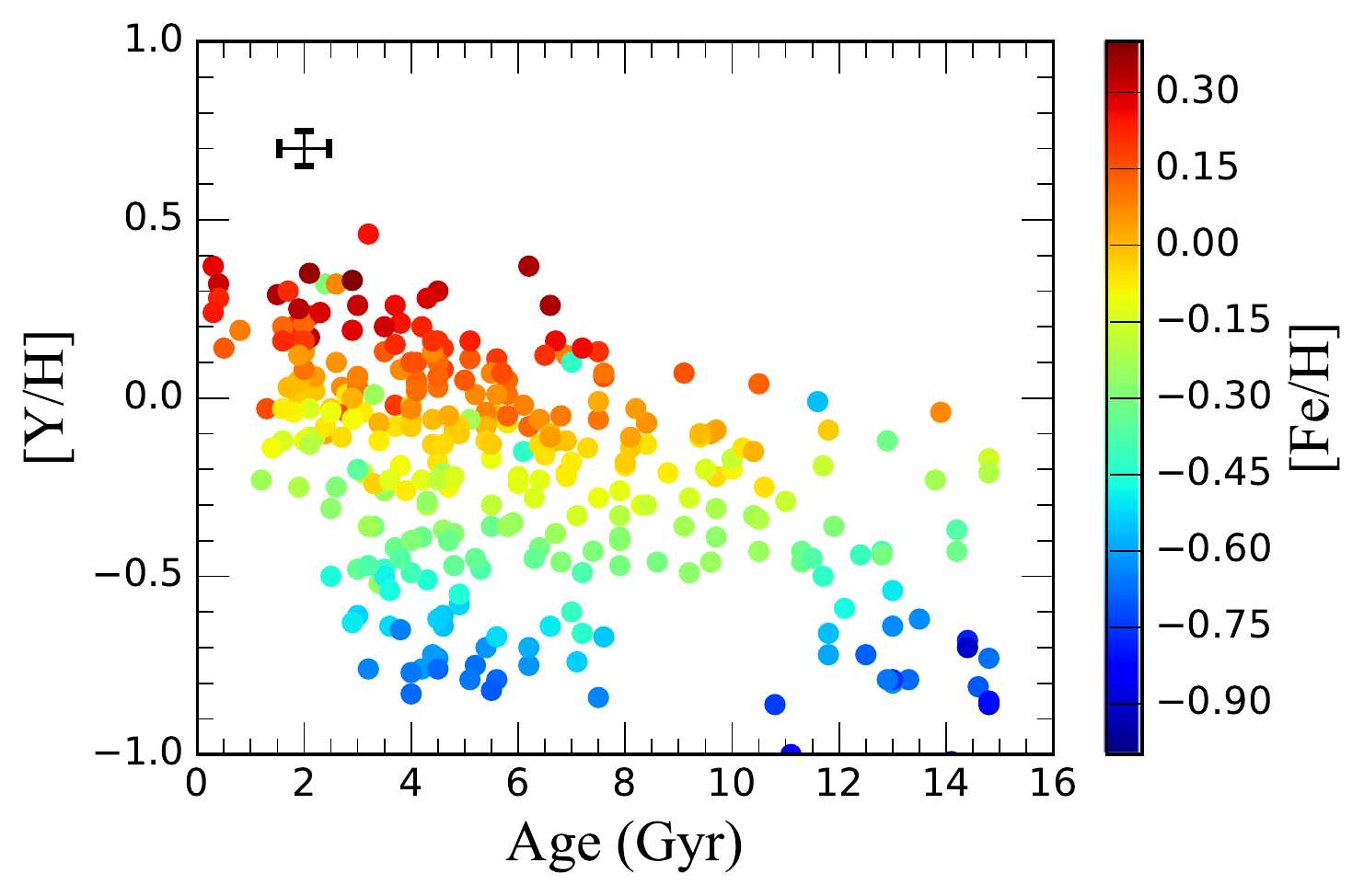}
  \vspace{-2mm}
      \caption{[Y/H] as a function of age (in Gyr) for all stars
      with $\sigma_{\rm Age}<$1\,Gyr from \citet{Bensby14}. The stars
      are colour coded according to their [Fe/H] (as indicated in the colour bar). A typical error bar is shown.}
         \label{fig:yage}
           \vspace{-5mm}
\end{figure}

%--------------------------------------------------------------------
\subsection{Origin of the [Y/Mg] -- age trend for solar twins}
\label{sect:investigation3.2}

The link between the star formation history of the Milky Way and the resultant Galactic chemical evolution has been well known for some time now \citep{Tinsley79,Matteucci03}, and the [Y/Mg] clock is a natural consequence of this. Y is a neutron-capture element produced through the $s$-process (or slow neutron capture process), which  occurs in asymptotic giant branch stars \citep[AGB stars, for a summary of nuclear reactions in AGB stars see][]{KandL14}.  All stars with masses between 1 and 8\,$M_{\odot}$ evolve to become AGB stars in their final phases. Due to this restricted mass range, the enrichment of the ISM with $s$-process material only begins roughly 500\,Myr after the formation of the first population of stars \citep[e.g.,][]{Sneden08}. Lower mass AGB stars (1 -- 4\,$M_{\odot}$) in particular produce large amounts of Y, meaning that the enrichment of the ISM with Y gradually increases with time \citep{Travaglio04,Fishlock14}, and therefore younger stars will have larger Y abundances (compare Fig.\,\ref{fig:yage}). 

Magnesium, in contrast, is an $\alpha$-element, and is mostly produced in Type\,II supernovae. These take place in the final stages of the evolution of massive stars (greater than 10\,$M_{\odot}$), and so typically occur  quickly after the stars formed. Therefore the ISM is almost immediately enriched in $\alpha$-elements, and becomes over-abundant in these compared to other elements. As time progresses and with the advent of Type Ia supernovae (which mainly produce
Fe, but little $\alpha$-element material) the relative $\alpha$-abundance compared to other elements such as iron is reduced \citep{Matteucci14}. Observationally, this trend is found in the Solar neighbourhood \citep{Fuhrmann98,Reddy03,Bensby03,Adibekyan12}, and on differing timescales in local dwarf spherodial galaxies \citep{Venn04}, and has been suggested as a useful `cosmic clock'  \citep[][see, however, Sect.\,\ref{sect:other}]{Gilmore89,Chiappini97,RecioBlanco14}.

Putting these two enrichment timescales together, and it is clear that young stars will have higher [Y/Mg] ratios than older stars, if all other variables are kept constant. This explains the trend seen by \citet{Nissen15}, \citet{Spina16}, and \cite{Tucci16}, where the stars examined had almost identical temperatures, surface gravities, and metallicities. In those works, it is suggested that [Y/Mg] can be used as an age indicator for these stars. However, once we start to extend this trend to other areas of parameter space, as we  do here, problems occur. Leaving aside stars of differing temperatures and surface gravities, it is clear that the evolution of both Y and Mg with time is not independent of the evolution of Fe (compare Fig.\,\ref{fig:all} and \ref{fig:yage}). Therefore once we start to vary the Fe abundance of the stars we examine, the trend breaks down.

\begin{figure}
  \centering
    \includegraphics[width=8.5cm]{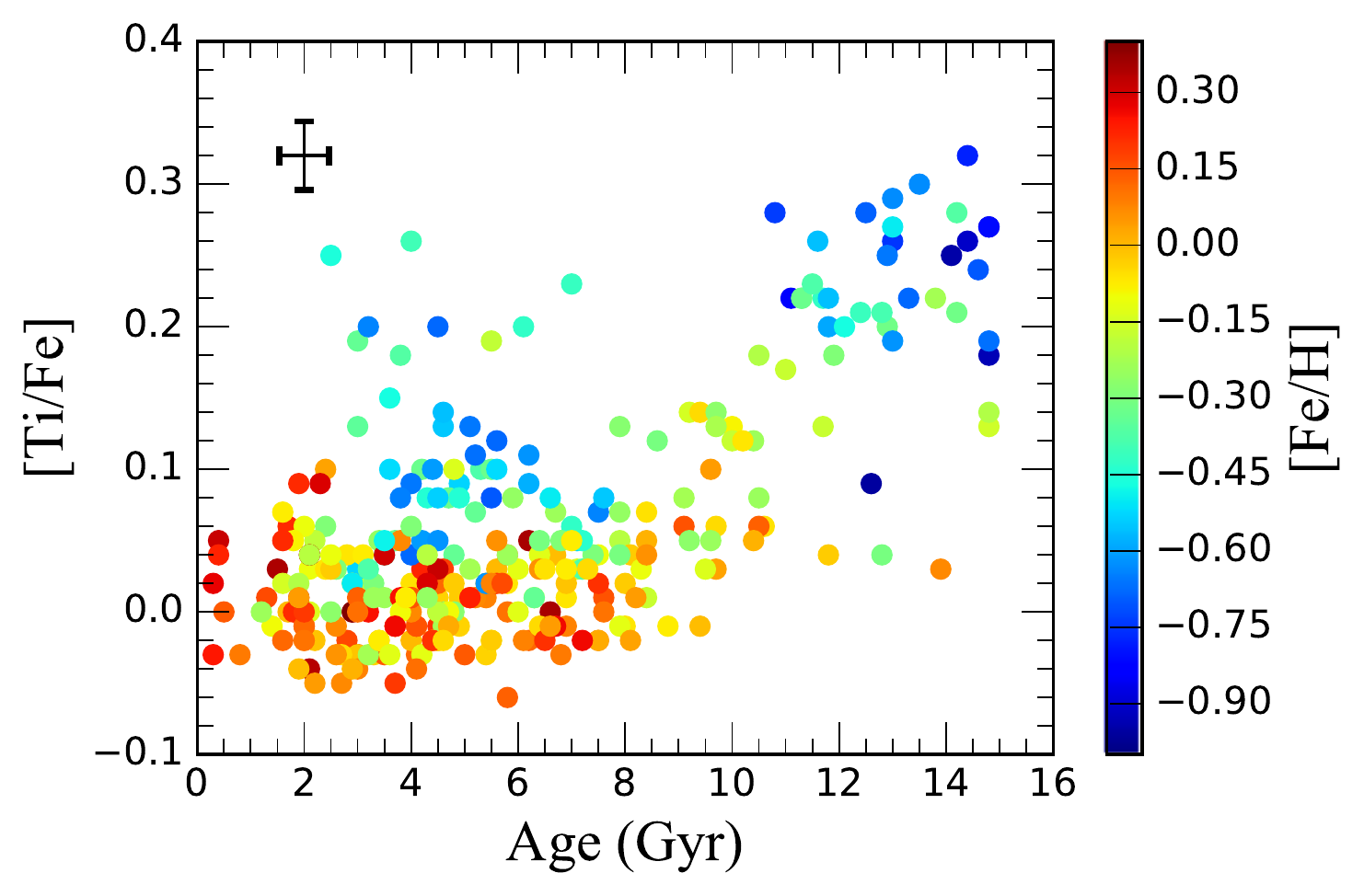}
  \vspace{-2mm}
          \caption{
          [Ti/Fe] as a function of age (in Gyr) for all stars
      with $\sigma_{\rm Age}<$1\,Gyr from \citet{Bensby14}. The stars
      are colour coded according to their [Fe/H] (as indicated in the colour bar). A typical error bar is shown. }
         \label{fig:tife}
           \vspace{-5mm}
\end{figure}

The amount of Fe in the ISM when a star forms will have a significant effect on the amount and type of $s$-process material that it produces \citep{KandL16}. Stars with a solar Fe abundance produce large quantities of elements formed close to the first $s$-process peak; Sr, Y, and Zr \citep{Travaglio04}. However, as the metallicity decreases, there are fewer Fe atoms per neutron to act as seeds for neutron-capture. This means less first peak elements are formed, instead greater abundances of second peak elements such as Ba and La form. This explains the strong gradient of [Fe/H] we see in Figure \ref{fig:yage} with regards to [Y/H].

If stars at lower Fe produce less Y, then the [Y/Mg] ratio would also be lower for the stars formed from the gas expelled - explaining why in Fig.\,\ref{fig:all} the ratio decreases with decreasing [Fe/H]. This means that, whilst more metal-rich stars have a negative [Y/Mg] trend with age, metal-poor stars are not enriched with enough Y to create this trend, and the age indicator no longer works.

We conclude that although it is relatively easy to find various abundance ratios that 
appear to vary in lock-step with age the interpretation of such relations is not obvious. There can
often be an underlying explanation from the point of view of nucleosynthesis or from
selection biases. A thorough understanding of both of these issues is required for a 
correct interpretation of the observations.

%--------------------------------------------------------------------
\subsection{Other  relations between elemental abundances and  age} 
\label{sect:other}

The usefulness of the $\alpha$-abundance `cosmic clock' as a proxy for age has been the subject of discussion in the literature \citep{Fuhrmann98,Ramirez13,RecioBlanco14}, and underpins the main argument for the [Y/Mg] clock as examined in this paper. Looking more specifically at this trend, Figure\,\ref{fig:tife} shows the relation between [Ti/Fe] and age. Similar figures have 
been presented in \citet{Bensby14} and \citet{Haywood13}. Here we 
explicitly indicate the [Fe/H] values for the stars as well as the Ti abundances and age
and find that the trend is not so straightforward, due to a large amount of [Fe/H]-related scatter in the plot. 
Figure\,7 in \citet{Haywood13} in fact shows a similar pattern, although perhaps
less obvious thanks to the colour coding scheme they use. We find that also the \citet{2016ApJS..225...32B} sample shows exactly the same pattern. At young ages, there are a wide range of [Ti/Fe] values seen, and in particular a significant group of more metal-poor stars with [Ti/Fe]$>0.1$. A star with [Ti/Fe]$=0.2$ and [Fe/H]$=-0.5$, for example, could have an age anywhere in the range of approximately 3 to 14\,Gyr. Stars with very different [$\alpha$/Fe] can have the same age, but highly varying [Fe/H]. Not only an observational effect, this is also seen in chemical evolution models such as that of \cite{RalphandPaul}.

The wide scatter in the trend, and the significant number of outliers, in the [$\alpha$/Fe] ratios cast some doubts on the usage of 
these ratios as indicators of age \citep[e.g., as applied in][]{Bovy12}.

%--------------------------------------------------------------------
\section{Conclusions}

We confirm the downward trend in [Y/Mg] with age first noted in {\cite{daSilva12} and thereafter explored further by \cite{Nissen15}. Our sample is fully independent in all aspects (spectra, spectral analysis, and age determination). Our sample has lower precision in the measurements (lower signal-to-noise ratios) but covers a wider range of types of stars and, crucially, a much larger range in [Fe/H].

We find that the downward trend strongly depends on the [Fe/H] of the stars. For solar [Fe/H] the trend is clear but for stars with [Fe/H] $\simeq$ --0.5\,dex, the trend is almost flat and has no predictive power on age. 
Our conclusion is thus that there exists clear a correlation between [Y/Mg] and age, but it 
depends on [Fe/H] too.  We find that independent 
samples show similar results and we have also tested our age determination and found that
regardless of the method used the results persist. 

We would propose the following investigations to fully conclude on the relation between [Y/Mg], age and [Fe/H]: The first `obvious' study would be a dedicated observing programme targetting stars with very similar stellar parameters as in, e.g., \cite{Nissen15}, but with [Fe/H] = --0.5\,dex. This should then be compared with the results from \citet{Nissen15}, \citet{Spina16} and \citet{Tucci16}. The best would be to select stars with effective temperature and surface gravities close to solar for a study that is as differential as possible between these studies and the new study. It would be exciting to perform similar studies also for stars {\it not} in the solar neighbourhood. Selection of suitable samples should be possible already with the upcoming TGAS data from the first {\it Gaia} release \citep{2015A&A...574A.115M}.

Another `obvious' study would be to go after stars on the red giant branch with asteroseismically determined stellar mass for which thus good ages can be derived and do a highly differential study of their elemental abundances both at solar metallcity and -0.5\,dex. This should show if the trends are also persistent for more evolved stars or not. Atomic diffusion might play a role when comparing turn-off stars and red giant stars, but it is expected that the differences should be smaller for higher metallicities \citep[see, e.g.,][]{2013A&A...555A..31G}.

\vspace{-3mm}
\section*{acknowledgments}

We thank Fan Liu for discussions about future tests of these trends.

S.F., L.M.H., P.J.M., and E.S. all acknowledge the grant The New Milky Way from the Knut and Alice Wallenberg Foundation. 
P.J.M. acknowledges support from the  Swedish National Space Board.
\vspace{-3mm}
\bibliographystyle{mnras}
\bibliography{references}

\bsp	% typesetting comment
\label{lastpage}
\end{document}